\documentclass[final,3p,times]{elsarticle}

\usepackage{amssymb}
\usepackage{xcolor}
\usepackage{amsmath}

\journal{Nuclear Instrumentation and Methods}

\begin{document}

\newcommand{\todo}[1]{\textcolor{red}{ToDO: #1}}

\begin{frontmatter}



\title{Performance measurements of the electromagnetic calorimeter and readout electronics system for the DarkQuest experiment}


\author[brandeis]{Aram Apyan}
\author[bu]{Christopher Cosby}
\author[ttu]{Yongbin Feng}
\author[bu]{Alp Gelgen}
\author[ucsc]{Stefania Gori}
\author[mit]{Philip Harris}
\author[jhu]{Xinlong Liu}
\author[purdue]{Mia Liu}
\author[jhu]{Petar Maksimovic}
\author[uva]{Cristina Mantilla-Suarez}
\author[bu]{Ryan McLaughlin}
\author[bu]{Catherine Miller}
\author[jhu]{Amitav Mitra}
\author[mit]{Noah Paladino}
\author[purdue]{Arghya Ranjan Das}
\author[ttu]{Valdis Slokenbergs}
\author[bu]{David Sperka}
\author[fnal]{Nhan Tran}
\author[bu]{Zijie Wan}

\affiliation[brandeis]{organization={Brandeis University Martin A. Fisher School of Physics},
            addressline={415 South Street}, 
            city={Waltham},
            postcode={02453}, 
            state={MA},
            country={U.S.A.}}

\affiliation[bu]{organization={Boston University Department of Physics},
            addressline={590 Commonwealth Avenue}, 
            city={Boston},
            postcode={02215}, 
            state={MA},
            country={U.S.A.}}

\affiliation[ttu]{organization={Texas Tech University Department of Physics},
            addressline={2500 Broadway W,}, 
            city={Lubbock},
            postcode={79409}, 
            state={TX},
            country={U.S.A.}}

\affiliation[ucsc]{organization={University of California Santa Cruz Department of Physics},
            addressline={550 Red Hill Rd, Santa Cruz, CA 95064}, 
            city={Santa Cruz},
            postcode={95064}, 
            state={CA},
            country={U.S.A.}}

\affiliation[mit]{organization={Massachusetts Institute of Technology Laboratory for Nuclear Science},
            addressline={77 Massachusetts Avenue}, 
            city={Cambridge},
            postcode={02139}, 
            state={MA},
            country={U.S.A.}}

\affiliation[jhu]{organization={Johns Hopkins University William H. Miller III Department of Physics and Astronomy},
            addressline={3400 N. Charles Street}, 
            city={Baltimore},
            postcode={21218}, 
            state={MD},
            country={U.S.A.}}

\affiliation[purdue]{organization={Purdue University Department of Physics and Astronomy},
            addressline={525 Northwestern Avenue}, 
            city={ West Lafayette},
            postcode={47907}, 
            state={IN},
            country={U.S.A.}}

\affiliation[uva]{organization={University of Virginia Department of Physics},
            addressline={382 McCormick Rd}, 
            city={Charlottesville},
            postcode={22903}, 
            state={VA},
            country={U.S.A.}}

\affiliation[fnal]{organization={Fermi National Accelerator Laboratory},
            addressline={PO Box 500}, 
            city={Batavia},
            postcode={60510}, 
            state={IL},
            country={U.S.A.}}

\begin{abstract}
This paper presents performance measurements of a new readout electronics system based on silicon photomultipliers for the PHENIX electromagnetic calorimeter. Installation of the lead-scintillator Shashlik style calorimeter into the SeaQuest/SpinQuest spectrometer has been proposed to broaden the experiment's dark sector search program, an upgrade known as DarkQuest. The calorimeter and electronics system were subjected to testing and calibration at the Fermilab Test Beam Facility. Detailed studies of the energy response and resolution, as well as particle identification capabilities, were performed. The background rate in the actual experimental environment was also examined. The system is found to be well-suited for a dark sector search program on the Fermilab 120 GeV proton beamline. 
\end{abstract}




\end{frontmatter}

\section{Introduction}
The DarkQuest experiment~\cite{Berlin:2018pwi,Apyan:2022tsd} is a planned upgrade of the SpinQuest experiment~\cite{Brown:2014sea,SeaQuest:2019hsx}, a proton fixed-target experiment at Fermilab. The SpinQuest experiment uses the SeaQuest spectrometer~\cite{SeaQuest:2017kjt} and is located in the NM4 experimental cavern at Fermilab. The goal of DarkQuest is to investigate dark sector particle decays to $e^+e^-$ final states via high-intensity proton interactions with an iron dump. The main feature of this upgrade is the inclusion of an electromagnetic calorimeter (EMCal), which is essential for precise measurements of the energy and position of electromagnetic showers, as well as for differentiating them from hadronic showers. A candidate EMCal detector~\cite{PHENIX:2003fvo} has been identified from the previous PHENIX experiment~\cite{PHENIX:2003nhg} at Brookhaven National Laboratory, thereby avoiding the cost of designing and building a new calorimeter. The PHENIX EMCal is a Shashlik-type electromagnetic calorimeter made out of alternating layers of lead and plastic scintillator. The total calorimeter is 18 radiation lengths and 0.85 nuclear interaction lengths in depth.  Detailed specifications of the EMCal can be found in Ref.~\cite{PHENIX:2003fvo}. The EMCal was successfully operated in the PHENIX experiment from 2000-2016 at the Relativistic Heavy Ion Collider. At PHENIX, it was extensively used for photon, pion, and electron identification for particle energies characteristic of DarkQuest ranging from 0.5~GeV up to 80~GeV.

However, the effects of potential radiation damage have not been studied to confirm that the detector still meets the experimental requirements for dark sector searches. Furthermore, the photomultiplier tubes (PMTs) used to detect the scintillation light require high power and are in unknown condition. A cost-effective solution is to replace the PMTs with silicon photomultipliers (SiPMs) mounted on a custom printed circuit board that can be read out and digitized using commercial off-the-shelf electronics. The studied commerical board is the CAEN DT5202 ASIC board, which is based on the WeeRoc\texttrademark~Citiroc ASIC. The performance of the detector, together with the associated readout electronics, was studied at the Fermilab Test Beam Facility~\cite{ftbf} during the summer of 2024. 

\section{EMCal Readout Electronics}

\begin{figure}[htbp]
    \centering
    \includegraphics[width=0.4\textwidth]{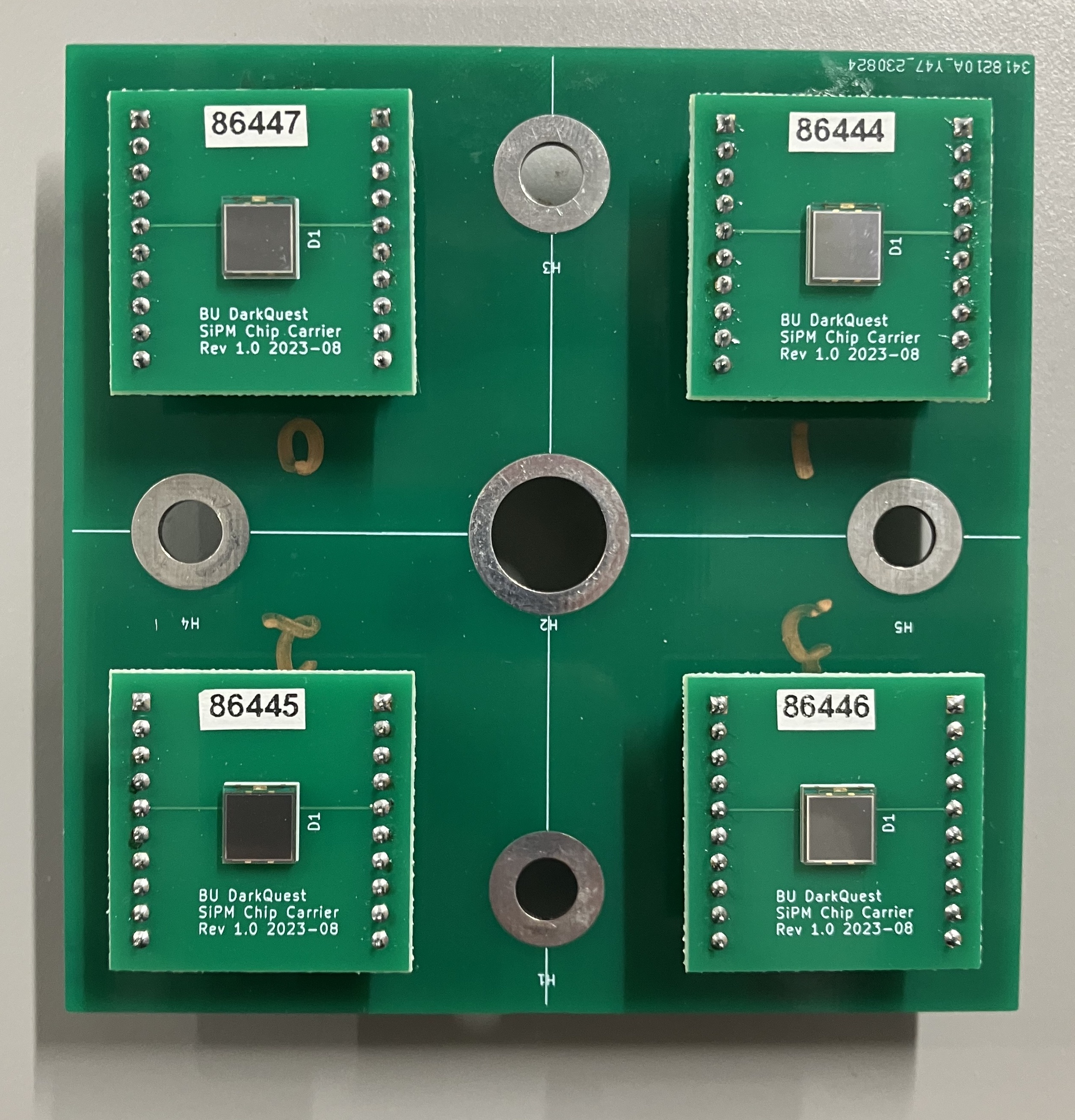}
    \includegraphics[width=0.56\textwidth]{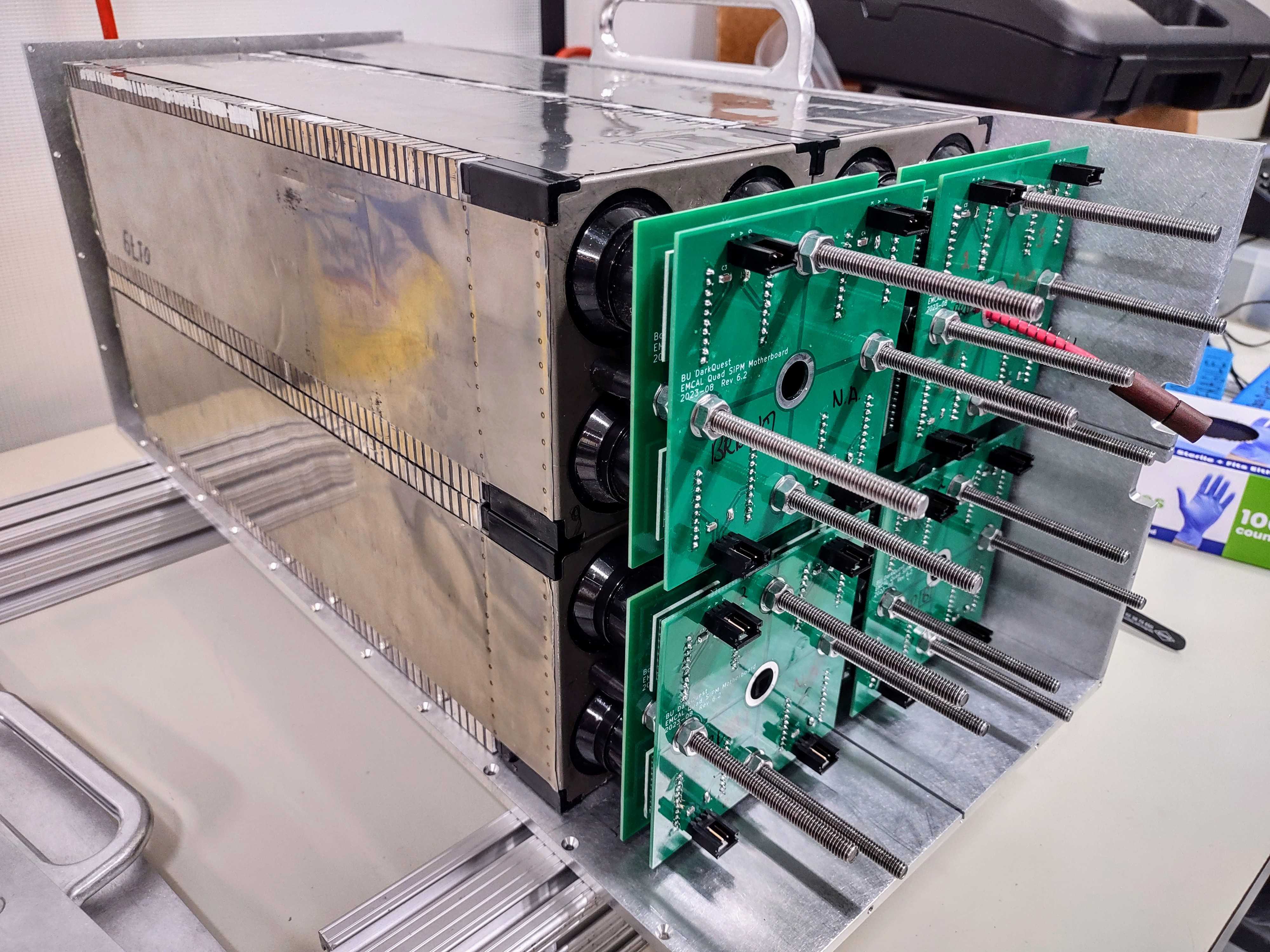}
        \caption{Left: A photograph of the 4 channel custom SiPM board. Right: A photograph of the 16 channel EMCal test module before being closed with the front end readout boards installed. }
    \label{fig:sipm-board-emcal-module}
\end{figure}

Each EMCal ``module'' consists of four independent readout channels, and the entire 2\,m x 4\,m EMCal sector consists of 648 modules (2592 total readout channels). Each channel corresponds to a 5.535 cm x 5.535 cm tower consisting of 66 layers of alternating lead and scintillator. A detailed illustration of a module can be found in Fig. 1 of Ref.~\cite{PHENIX:2003fvo}. The light of each channel is collected by through-going optical fibers concentrated into a 6mm diameter bundle of 36 optical fibers, which is well matched for a single 6\,mm x 6\,mm SiPM. These optical fiber ends were  repolished after removing the old PMTs. While there is a small gap between the SiPMs and the optical fibers, the use of 4 alignment studs as well as a foam ring around the fiber ends allows for each SiPM to be exactly aligned with the fiber ends. This gap is approximately one millimeter. The EMCal is expected to produce $\mathcal{O}(10,000)$ photons at the optical fiber ends per GeV of energy. Tests of the SiPM response conducted prior to the test beam showed a close match between the expected and observed SiPM saturation curves, suggesting the photon loss between the fiber ends and the SiPM surface is not significant.
The selected SiPM is the Hamamatsu\texttrademark~S14160-6010PS, which has a photon detection efficiency of approximately 18\% at the peak of the EMCal's emission spectra ($\approx$490nm) and an intrinsic gain of $\sim \! 10^5$. Furthermore, these SiPMs have a 10\,$\mu$m pixel pitch and contain 359011 pixels, which helps prevent saturation at the highest energies. Other SiPM models were tested prior to the use of the test beam and found to saturate at lower energies. Validating the suitability of the selected SiPM was a primary goal of the test beam studies. Custom front-end electronics boards that house the SiPMs and send the analog signal off the detector were developed and can be seen in Fig.~\ref{fig:sipm-board-emcal-module} (left). The board consists of a chip-carrier onto which the SiPM is reflow soldered, and a separate mother board which contains the other circuit components. Each channel has an individual cable connection to the CAEN DT5202. The modular design allows for easy switching between different SiPM models or testing different readout circuits. 

The target dynamic range of the EMCal is between energy deposits of 250 MeV and 80 GeV. During tests with cosmic rays it was predicted that some level of attenuation of the detector signal would be necessary to prevent saturation of the SiPM signal or the DT5202 circuit.  Therefore, a primary goal of the beam tests were to study the effects of optical and electrical attenuation on the performance of the EMCal. For the purpose of optical attenuation, we used a 1-stop neutral density filter attached to a frame placed between the optical fibers and the SiPMs. In order to test electrical attenuation, the cables connecting the SiPMs to the DT5202 were spliced with an SMA connection. This allows for the ability to switch between various electrical attenuation options without disturbing the EMCal. In particular, we chose to test a 10dB electrical attenuator. The expected effects from both sources of attenuation were tested prior to their usage in the test beam using studies of the SiPM response to direct LED light as well as with cosmic muons.

\section{Experimental Setup}
\begin{figure}[!htbp]
    \centering
    \includegraphics[width=0.8\textwidth]{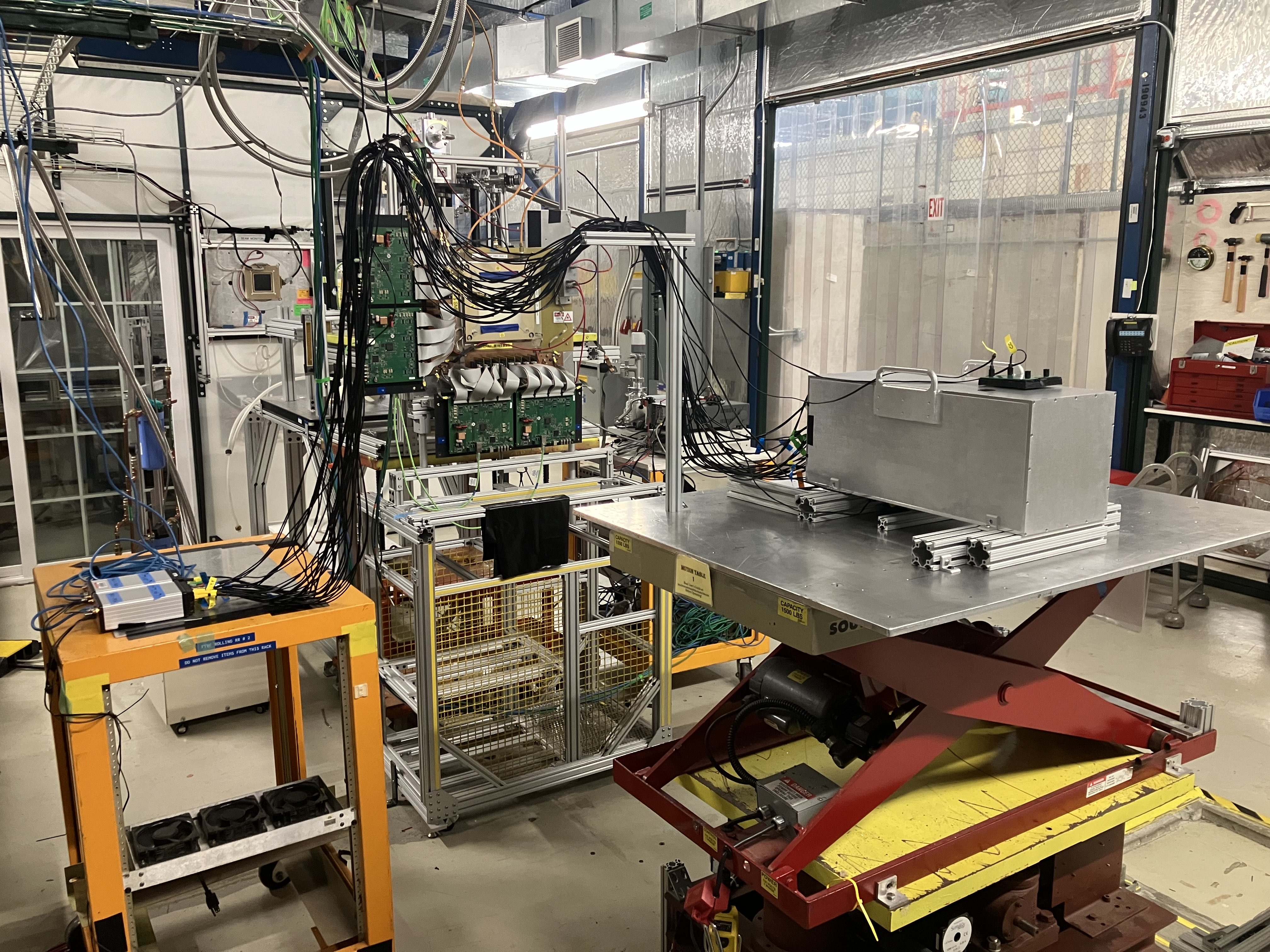}
        \caption{A photograph of the 16 channel EMCal test module installed on the test beam motion table.}
    \label{fig:emcal-test-beam-setup}
\end{figure}

The experiment was located on the MTest beamline at the Fermilab Test Beam Facility~\cite{ftbf}. This beamline is designed for testing detectors and has several beam configurations capable of providing protons, pions, electrons, and muons in the range of 1 to 120 GeV/c.  It can provide rates in the range of 1,000 to 300,000 particles per spill, each spill lasting 4 seconds, and with rates monitored by Cherenkov and scintillation detectors. The muon beam is achieved using an upstream beam absorber in the low-energy pion beam mode and yields several thousand muons per spill over a 1 m x 1 m area. The other beam modes provide beams with a spot size ranging from 6 mm to 13 mm. A 16 channel EMCal test module was built by placing 4 EMCal modules in an aluminum dark box as seen in Fig.~\ref{fig:emcal-test-beam-setup}. The test module was set up to trigger on energies above a predetermined threshold. A photograph of the test module before being closed can be seen in Fig.~\ref{fig:sipm-board-emcal-module} (right). The module was placed on a motion table and aligned with the beam using a laser system. The DT5202 was placed roughly 2\,m away from the EMCal and thus well outside the beamline. Additional scans of the motion table position were performed with the beam to study the detector alignment. The momentum spread of the beam was not independently determined, but has previously been reported to be between 2\% and 3\%~\cite{sPHENIX:2021ieee} and is expected to be worse for lower energy particles. 

The data collected for our purposes swept the energy range of the MTest beam in the pion configuration and in the muon configuration. Aside from the high-intensity tests, the beam intensity was kept at 10k particles per spill to prevent pile-up buckets. The accuracy of the spatial calibration was verified at the start of testing by remotely moving the motion table in a grid pattern between spills.

\section{Calibration with muons}
As minimum ionizing particles (MIP), muons provide a straightforward calibration method. For muons that traverse longitudinally through an entire channel, the expected energy deposition can be calculated from the Bethe-Bloch equation, and the most probable value is about 235 MeV. At DarkQuest, muons are copiously produced in the experimental cavern as secondary decay products of hadrons  formed when the 120 GeV proton beam is dumped into a 5m-long iron magnet (FMag).

\begin{figure}[htbp]
\centering
\includegraphics[width=0.4\textwidth]{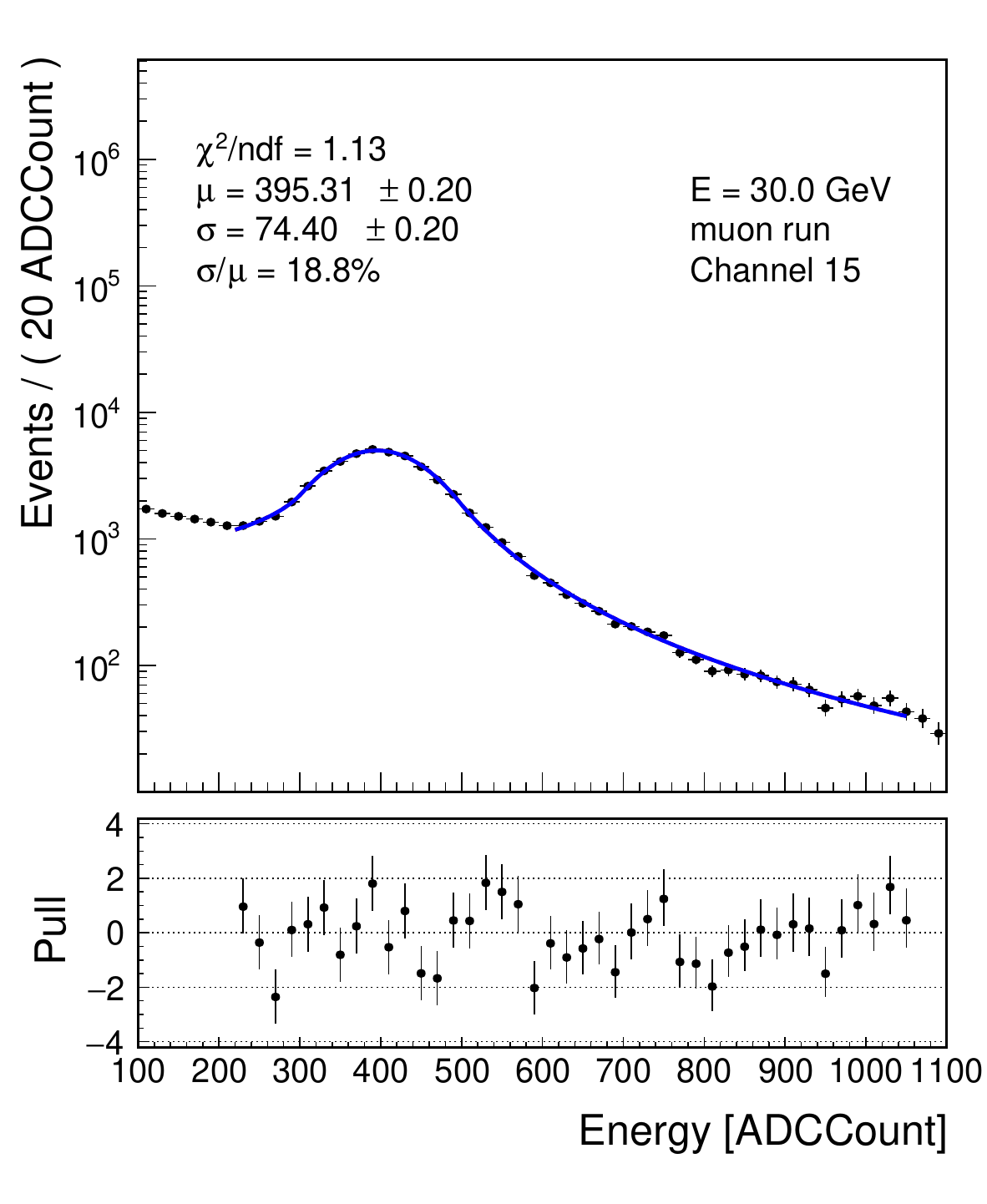}
\caption{Example of a fit with a double-sided Crystal Ball function on a channel for a muon run.}
\label{fig:crystal_ball}
\end{figure}

Energy deposits in the EMCal during muon runs of the MTest beam were analyzed channel-by-channel, and each channel was fitted with a double-sided crystal ball function~\cite{Gaiser:1982yw,Oreglia:1980cs}. An example fit is shown in Fig.~\ref{fig:crystal_ball}. Similar calibrations were performed for the runs with neutral density filters or electrical attenuators.  The double-sided crystal ball function takes on the general form:

\[
f(m; m_0, \sigma, \alpha_L, n_L, \alpha_R, n_R) =
\begin{cases}
  A_L \cdot \left( B_L - \frac{m - m_0}{\sigma_L} \right)^{-n_L}, & \text{for } \frac{m - m_0}{\sigma_L} < -\alpha_L \\
  \exp \left( - \frac{1}{2} \cdot \left[ \frac{m - m_0}{\sigma_L} \right]^2 \right), & \text{for } -\alpha_L\leq\frac{m - m_0}{\sigma_L} < 0 \\
  \exp \left( - \frac{1}{2} \cdot \left[ \frac{m - m_0}{\sigma_R} \right]^2 \right), & \text{for } 0\leq \frac{m - m_0}{\sigma_R} < \alpha_R \\
  A_R \cdot \left( B_R + \frac{m - m_0}{\sigma_R} \right)^{-n_R}, & \text{for } \alpha_R\leq \frac{m - m_0}{\sigma_R}.
\end{cases}
\]

where \( A_L \) and \( B_L \) are defined by:

\begin{align}
    A_i &= \left( \frac{n_i}{|\alpha_i|} \right)^{n_i} \cdot \exp\left( - \frac{|\alpha_i|^2}{2} \right), \\
    B_i &= \frac{n_i}{|\alpha_i|} - |\alpha_i|.
\end{align}

and $\alpha$ and $n$ correspond to the threshold values for change in behavior and the powers in the tails, respectively.

\begin{figure}[htbp]
\centering
\includegraphics[width=0.48\textwidth]{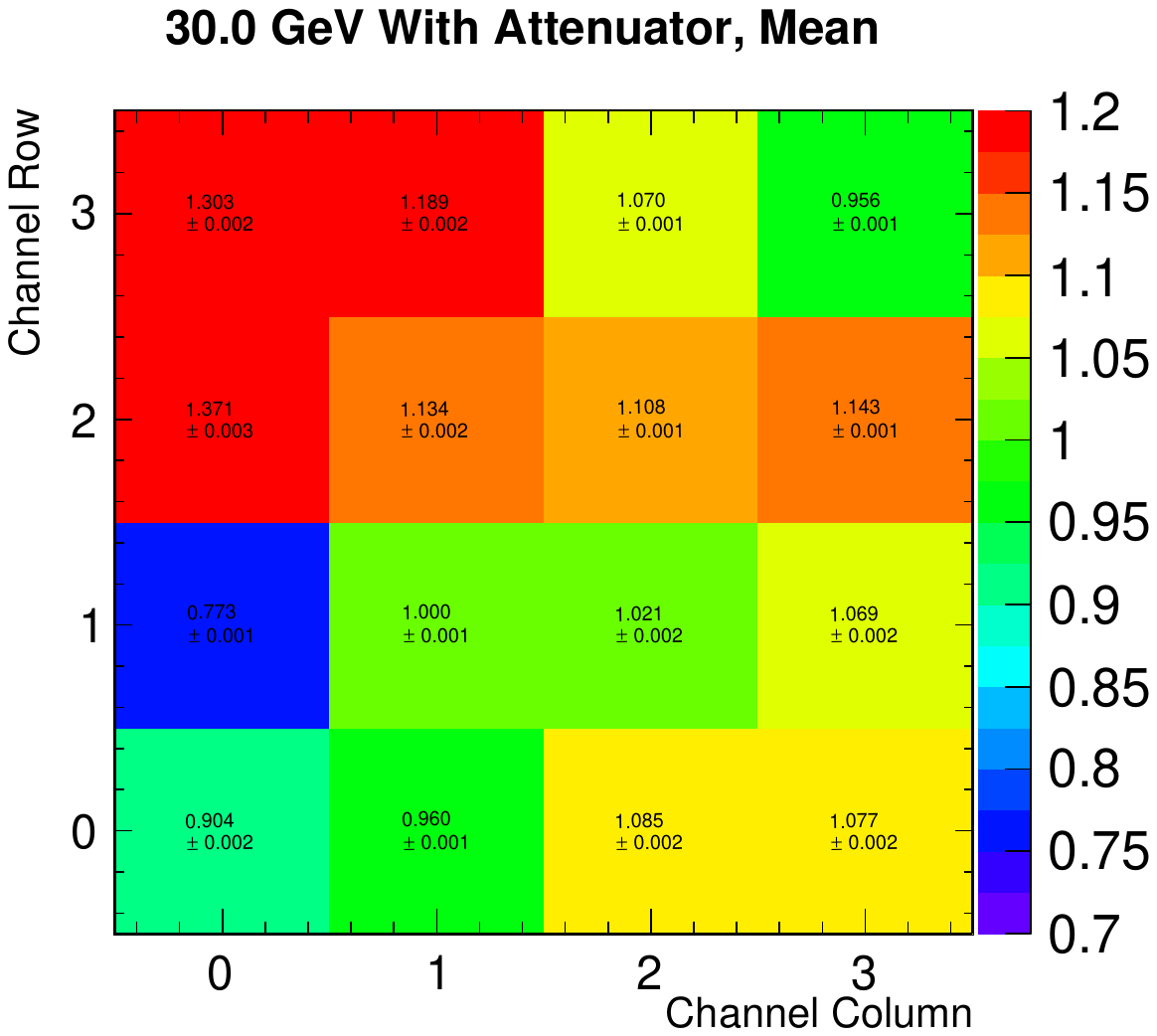}
\includegraphics[width=0.48\textwidth]{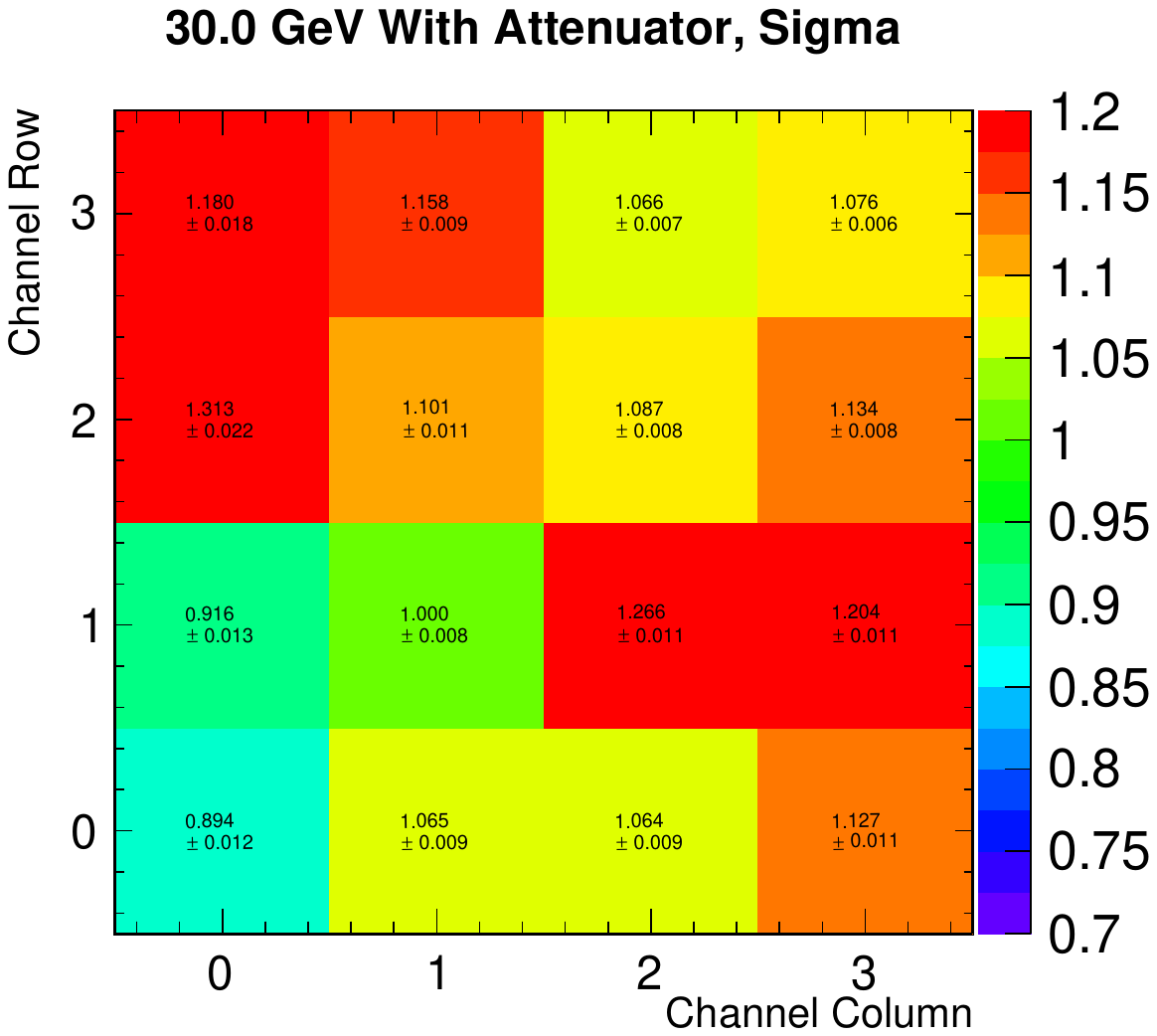}
\includegraphics[width=0.48\textwidth]{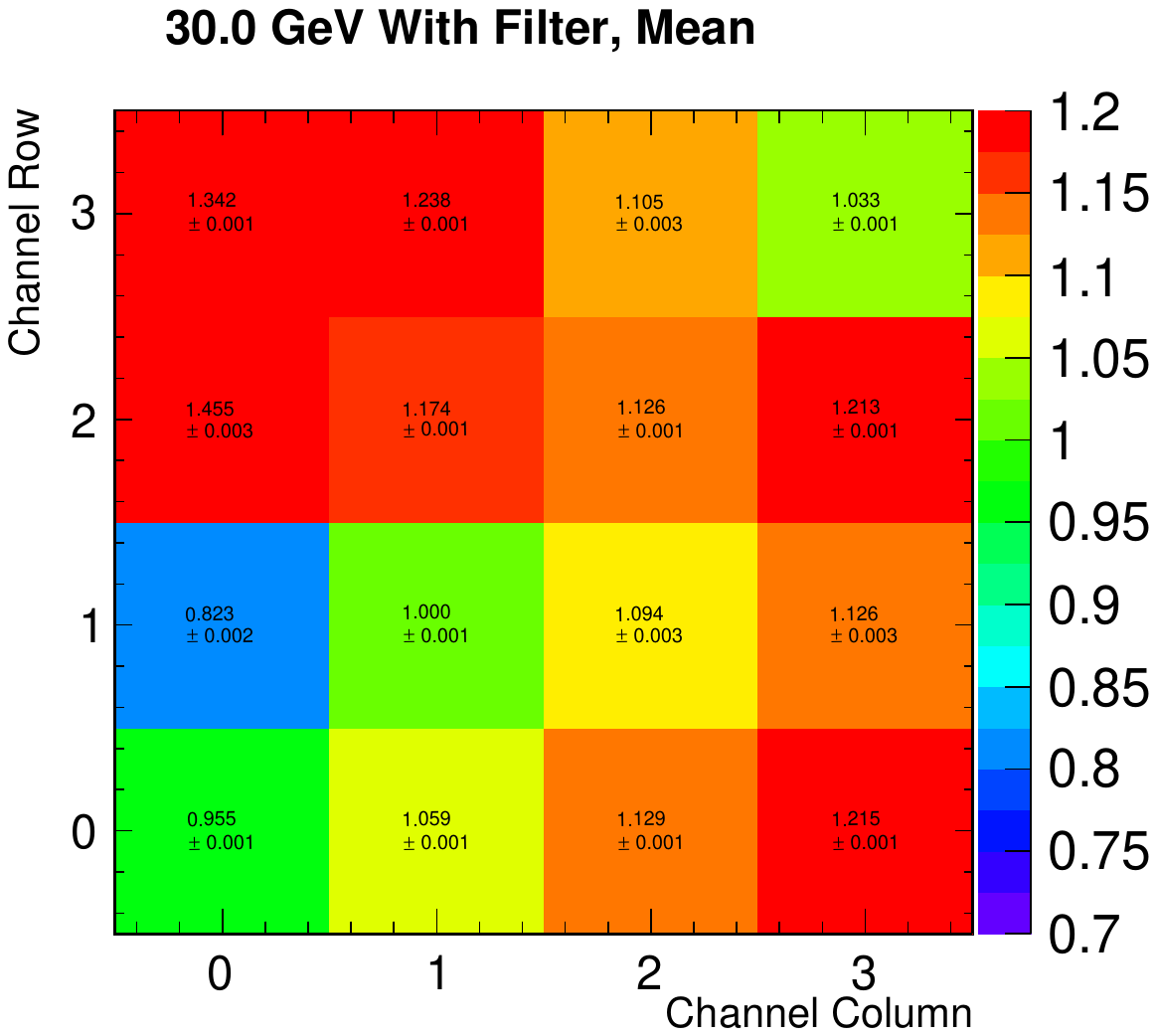}
\includegraphics[width=0.48\textwidth]{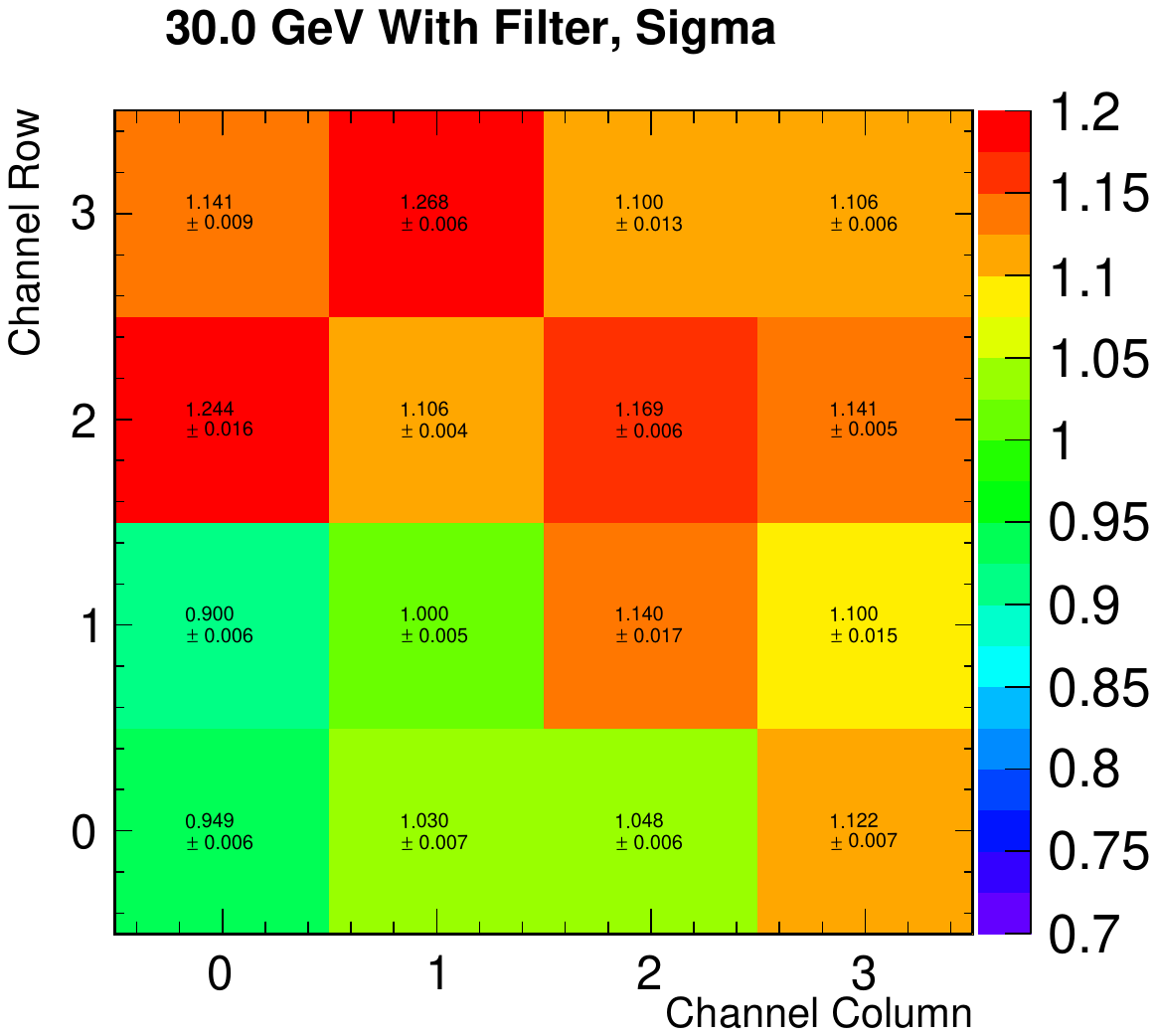}
\caption{Mean (left) and sigma (right) from the MIP fits with a double-sided crystal Ball function, using the runs with attenuators (top) and neutral density filters (bottom). The 30 GeV refers to the energy of the primary beam, the secondary muons have a broad range of momenta.}
    \label{fig:mipcalib_results}
\end{figure}

The mean ($m_0$) and sigma ($\sigma$) of these fits for different EMCal channels with attenuators and with neutral density filters are shown in Fig.~\ref{fig:mipcalib_results}. Between the two configurations, both mean and sigma distributions look similar with some variation. When comparing the attenuator distribution to the filter distribution, there is an average channel by channel difference of 5.8\% for the means and 4.3\% for the sigmas. This difference is attributed to the different distance between the SiPM and the fiber bundle before and after adding the neutral density filters. The energy distribution may also have been affected by the tightness of the cable connection when adding or removing the SMA attenuators. The variation of the calibration factors between different channels is around 30\%. This is a large spread and is indicative of an aged, and in some cases damaged, detector. In particular, one of the 4-channel modules that was used for lab testing sustained damage to some of the optical fibers during shipping and clearly shows a lower response than the other modules. 

\section{EMCal Detector Performance}

\begin{figure}[htbp]
\includegraphics[width=0.33\textwidth]{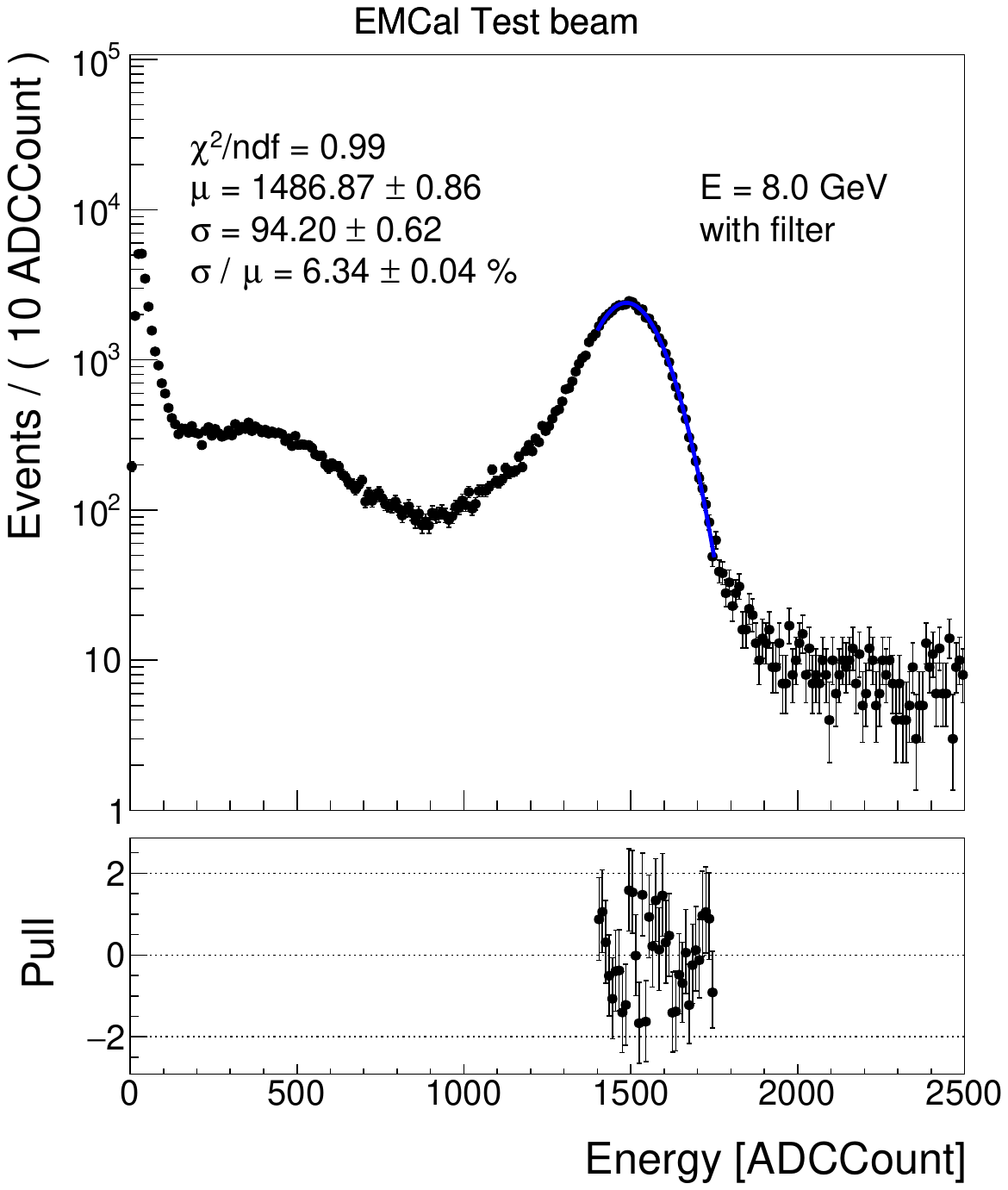}
\includegraphics[width=0.33\textwidth]{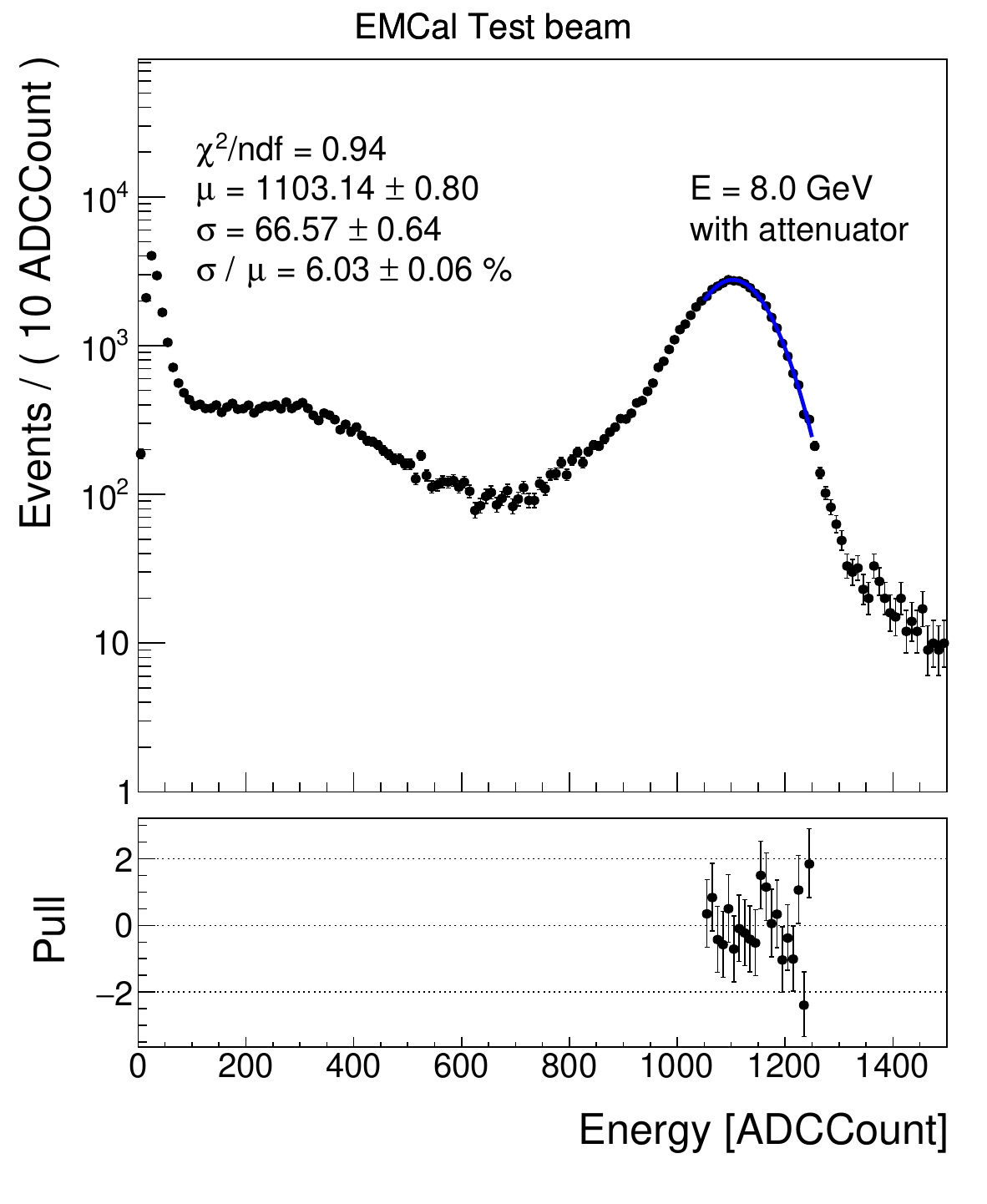}
\includegraphics[width=0.33\textwidth]{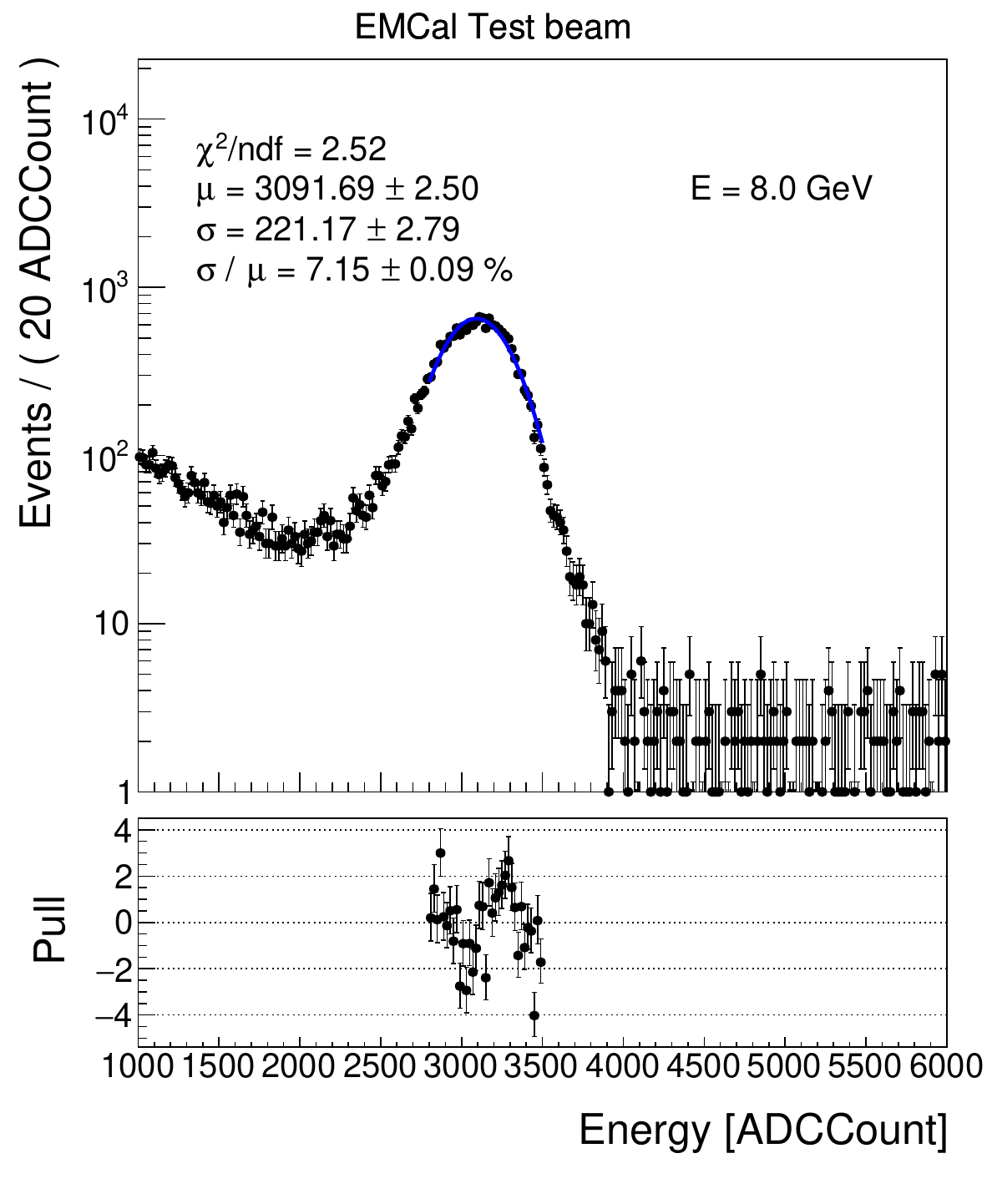}

\caption{Example fits to extract the EMCal response and resolution, fitted to a Gaussian function in the peak region.}\label{fig:emcal-fit}
\end{figure}

The EMCal's linearity and resolution as functions of energy are key parameters in understanding the performance of the calorimeter and ensuring its reliability in the DarkQuest experiment. These parameters were studied using the electrons present in low-energy pion runs, where, depending on the beam energy, between 10\% to 50\% of the beam is expected to be composed of electrons and the remaining particles mostly being pions.  The ADC values in the EMCal are corrected channel-by-channel and summed together after MIP calibration. The calibrated and summed energy distribution is then fit with a Gaussian function in the peak region, which corresponds to the energy deposit of the incoming electron beam. By varying bounds until the reduced chi-square exceeds 1.5, our fitting procedure gives an additional mean uncertainty of 0.5 ADC and for sigma 0.2 ADC. This is less than the statistical uncertainty. The pion contribution, concentrated at lower ADC values, is suppressed by two orders of magnitude in the peak region. The noise peak near zero is also well separated from the full-energy peak (see Fig.~\ref{fig:emcal-fit}). 

\begin{figure}[htbp]
\includegraphics[width=0.5\textwidth]{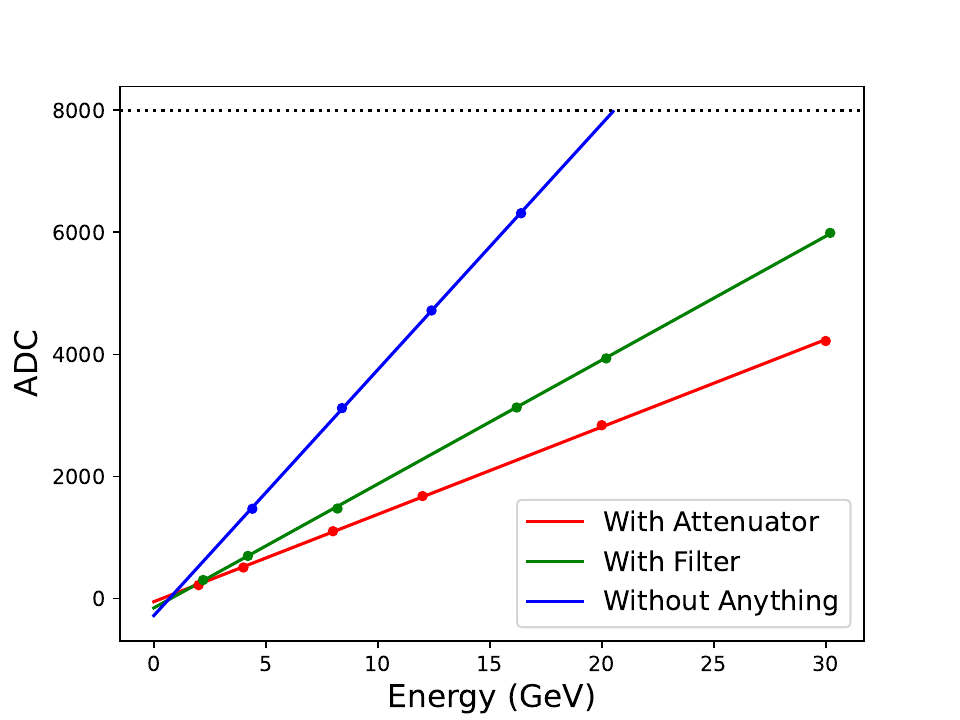}
\includegraphics[width=0.5\textwidth]{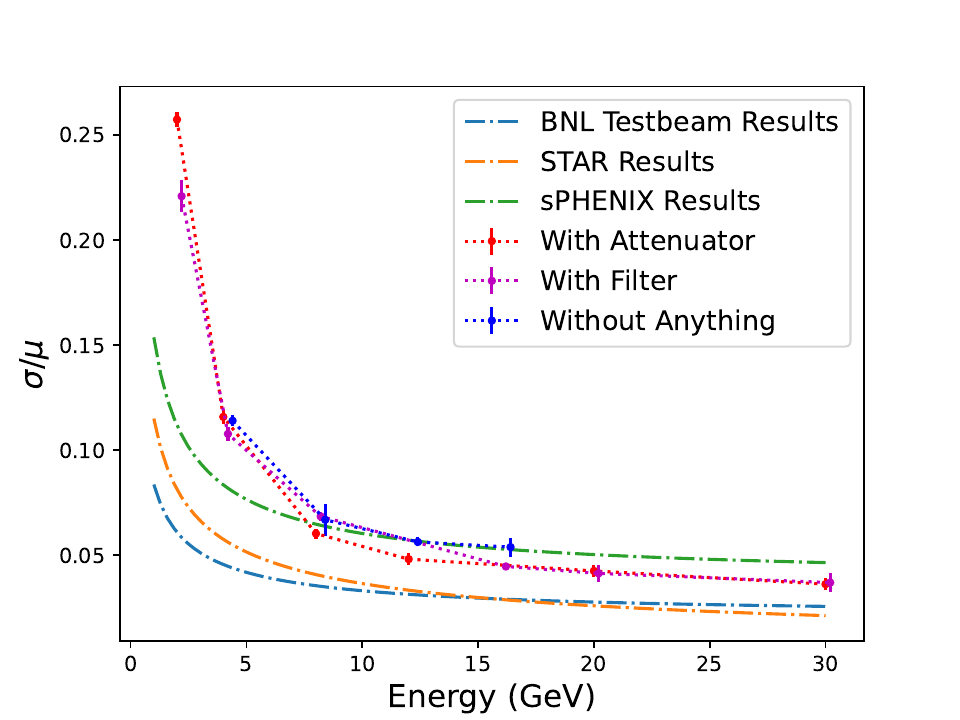}
\caption{Left: EMCal response as a function of beam energy. Note the maximum readout ADC for a single channel is 8000. Right: EMCal resolution as a function of beam energy compared with other similar experiments~\cite{sPHENIX:2021ieee,bnl:resol,star:resol}. Note the ``with filter" data points have been shifted 0.2 GeV to the right and ``without anything" data points have been shifted 0.4 GeV to the right. This was done to allow each point to be clearly visible. 
}\label{fig:emcal-resol}
\end{figure}

The fits were carried out for different incoming beam energies, varying from 2 GeV to 30 GeV. The energy response (mean) and resolution are extracted from the fits. These are shown as a function of the beam energy in Fig.~\ref{fig:emcal-resol}. In all the cases, the response distribution is found to be reasonably linear up to 20 GeV. However, beyond this value the charge delivered to the DT5202 is beyond its dynamic range. As we are interested in particle energies up to 80 GeV, we tested two different methods of attenuation, optical or electrical, to gain sensitivity to higher energies. Both forms of attenuation demonstrated a linear response for all energies tested. 

The EMCal performed well for our purposes. A comparison of our measurement with measurements from previous experiments for the same detector is shown in Fig.~\ref{fig:emcal-resol}. While the energy resolution is not as good as what was measured previously~\cite{bnl:resol}, it performed on par with, and in some cases better than, what was found with other calorimeters such as the sPHENIX calorimeter~\cite{sPHENIX:2021ieee}. This is especially true for particle energies above $\sim \! 10$ GeV, which we expect to be satisfied by $\sim\!90$\% of signal events.  As discussed earlier, the decrease in the detector's performance when compared to previous experiments is likely twofold in origin: imperfections in the experimental environment and degradation with age. 
In particular, we encountered complications related to the durability of the cable connections to the SiPM boards. This led to the connections becoming loose as the detector housing was being closed. For this reason, the boards and cables will be redesigned for increased durability using more secure connectors (e.g. MMCX). However, because of the uncertainty in the MTest beam momentum spread, the detector's actual resolution may be better than what we observed. Beyond these effects, this detector is around 20 years old and has sustained significant irradiation, causing some degradation to the scintillating plastic performance over the years.

\begin{figure}[htbp]
    \centering
        \includegraphics[width=0.5\textwidth]{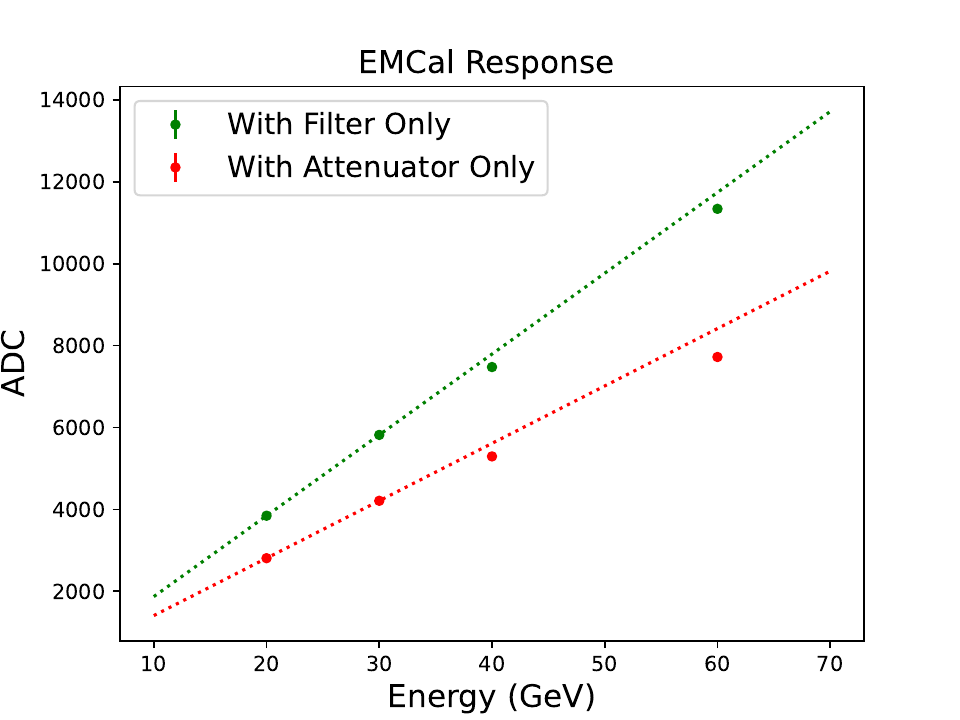}
    \caption{ADC as a function of the beam energy for data taken with an electrical attenuator only and with an optical filter only. These runs had an intensity of roughly 200k counts per spill and were focused at the center of an individual channel. Note that the lines of fit shown are linear fits from energies less than or equal to 30 GeV extrapolated to higher energies in order to gauge non-linearity.}
    \label{fig:energy_resol_to_80}
\end{figure}

Particles produced by the low-energy pion beam have a maximum energy of 30 GeV. In an attempt to test the linearity beyond this threshold, several runs with high intensity ($\mathcal{O}(10^5)$ particles per spill) were taken. The purpose of the high intensity runs was to increase the probability that two adjacent RF buckets both contain a particle, or even ``pile-up" buckets, in which a single bucket contains two particles. In these cases, the combined energy of the two particles was measured by the EMCal. Since the SiPM recovery time is on the order of 100 ns and the electronic integration time is on the order of 10 $\mu\text{s}$, these high-intensity runs should be largely indistinguishable from the standard runs. As shown in Fig.~\ref{fig:energy_resol_to_80}, we were able to successfully test the linearity with both forms of attenuation for energies up to 60 GeV and saw a lower relative non-linearity when using the optical filter than when using the electrical attenuator. If the beam was directed between two channels, the response remained linear as distributing the energy between the channels ensures the individual readings remain low. However, if the beam was instead directed directly into the center of one channel, a non-linearity of $\sim\!8$\% was observed. For the purpose of achieving a linear response, optical attenuation is expected to be more effective than electrical attenuation. This is most immediately due to optical saturation of the SiPMs for high energies showers, as these can produce approximately the same number of photons as the number of pixels in the SiPM. The effect of non-linear response on the particle identification performance (PID) was studied using simulations of electrons and pions with energy between 10 and 70 GeV. The effect was found to be marginal across the full energy spectrum until the non-linearity becomes larger than about 30\%. Therefore it is determined that both attenuation methods result in adequate performance.

\section{Electron and pion separation}

The shower shape of particles in the EMCal in otherwise similar events could be a useful tool for improved PID. This section details the differences in shower shape between electrons and pions in the EMCal, confirms this effect is an effect of particle flavor, not energy, and demonstrates the PID performance using only the calorimeter. The composition of the test beam used in this study is dependent on the energy of the test beam, but ranges from 50\% electrons down to 10\% electrons as the energy increases~\cite{ftbf}. Importantly, pions, unlike electrons, do not tend to deposit their entire energies in the detector as the EMCal is less than one nuclear interaction length in depth. As such, the energy deposited by electrons is typically higher than that of pions.

\begin{figure}[htbp]
\centering
\includegraphics[width=0.45\textwidth]{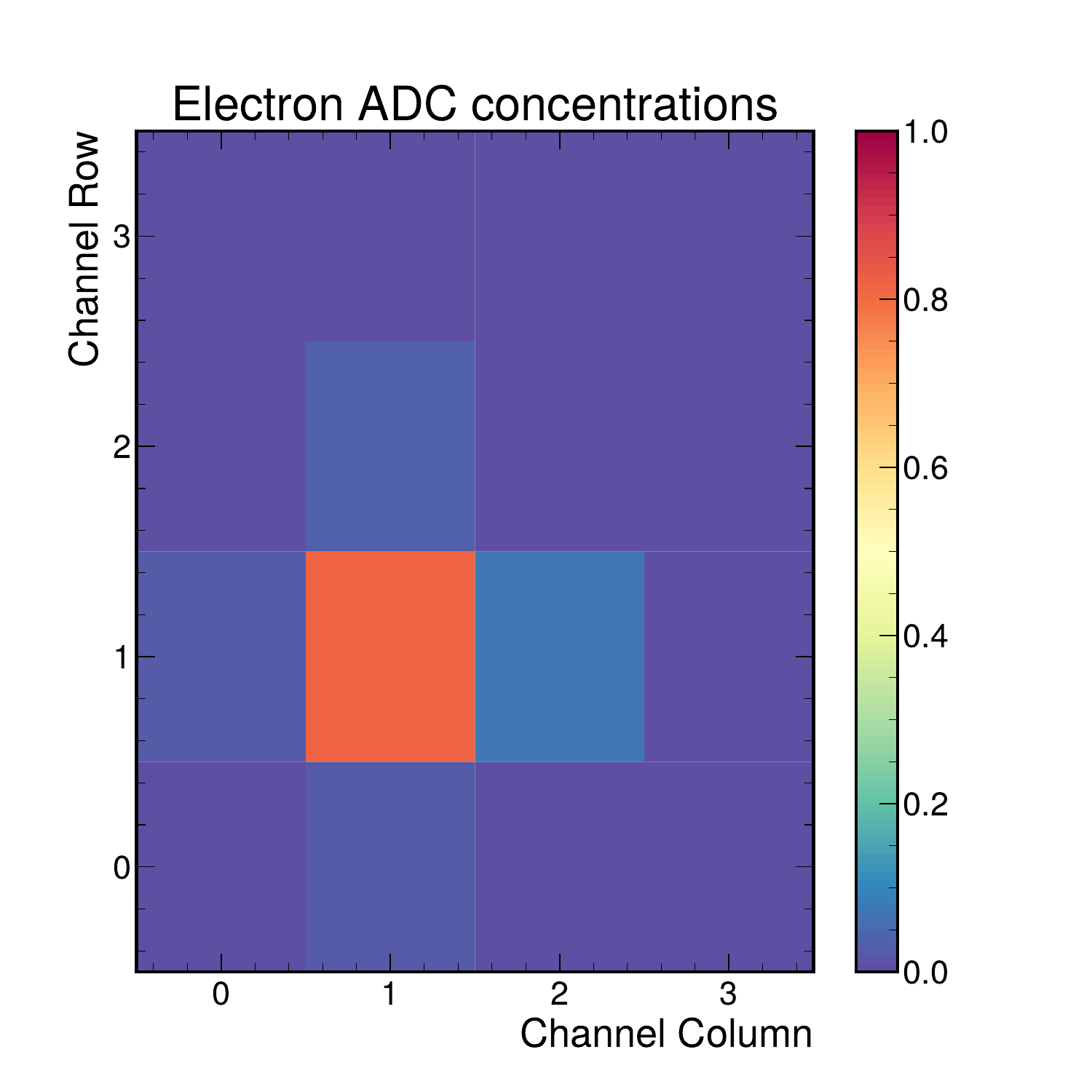}
\includegraphics[width=0.45\textwidth]{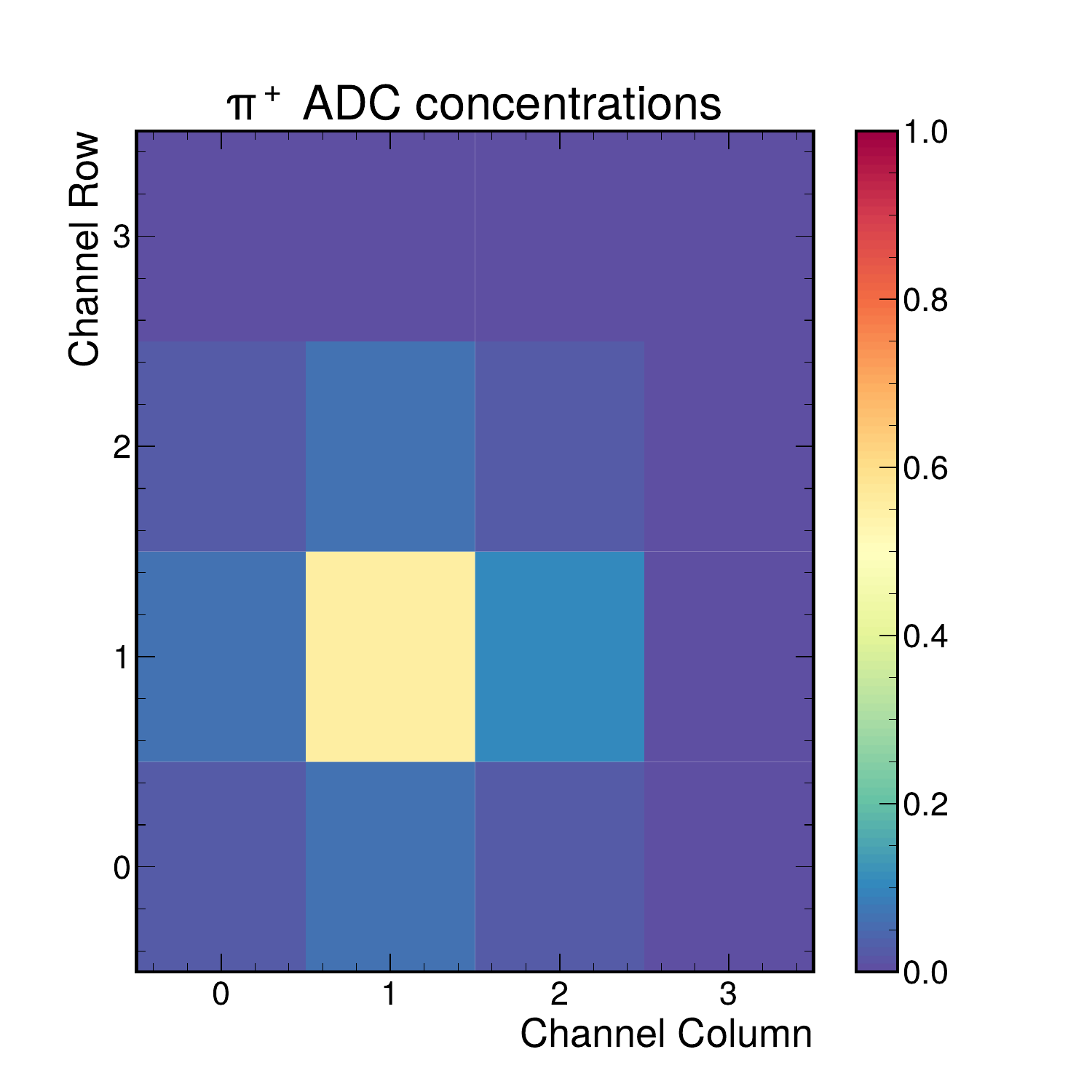}
\caption{Energy fraction per channel in a typical 16 GeV run after selecting electron (left) and pion (right) regions from summed ADC distribution. }\label{fig:2d_histos}
\end{figure}

The distribution of the summed ADC count per event for an entire run (e.g. Fig.~\ref{fig:emcal-fit}) can be used to identify approximately pure electron and pion regions. A narrow cut around the Gaussian peak is taken to identify an electron region, and the plateau between the noise peak (close to 0) and the Gaussian is dubbed our pion region. In the following, we only consider runs in which the majority of the energy is deposited in one of the four middle channels. The study is performed across three beam energies per test beam configuration (i.e. with neutral density filter or with attenuator). With these aforementioned selections, it is possible to assess the differences between electrons and pions. The distribution of the fractional energy per channel for electrons and pions is shown in Fig.~\ref{fig:2d_histos}. A clear difference between the spectra is observed.  

\begin{figure}[htbp]
\centering
\includegraphics[width=0.32\textwidth]{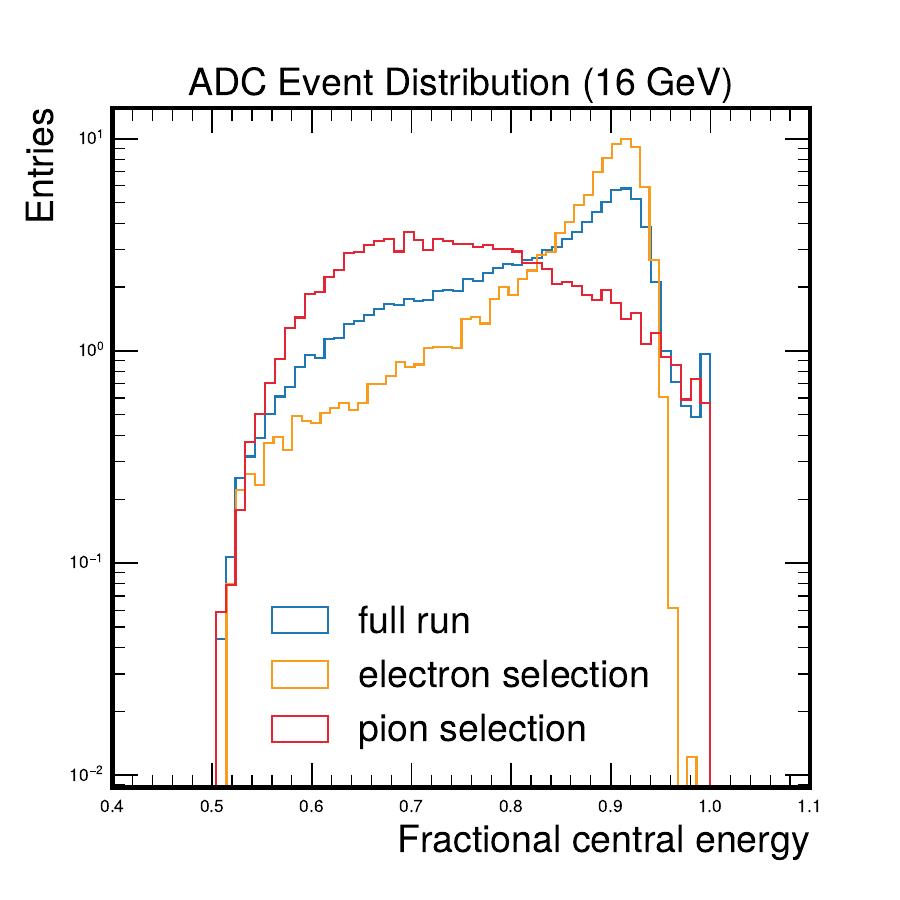}
\includegraphics[width=0.32\textwidth]{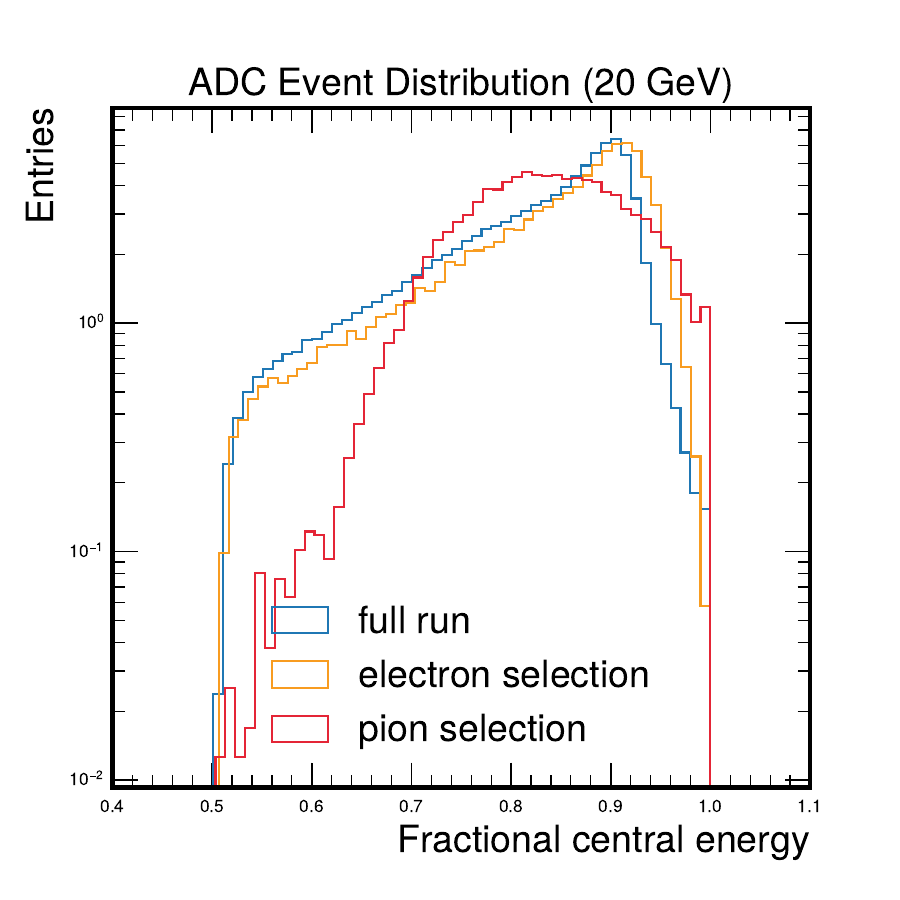}
\includegraphics[width=0.32\textwidth]{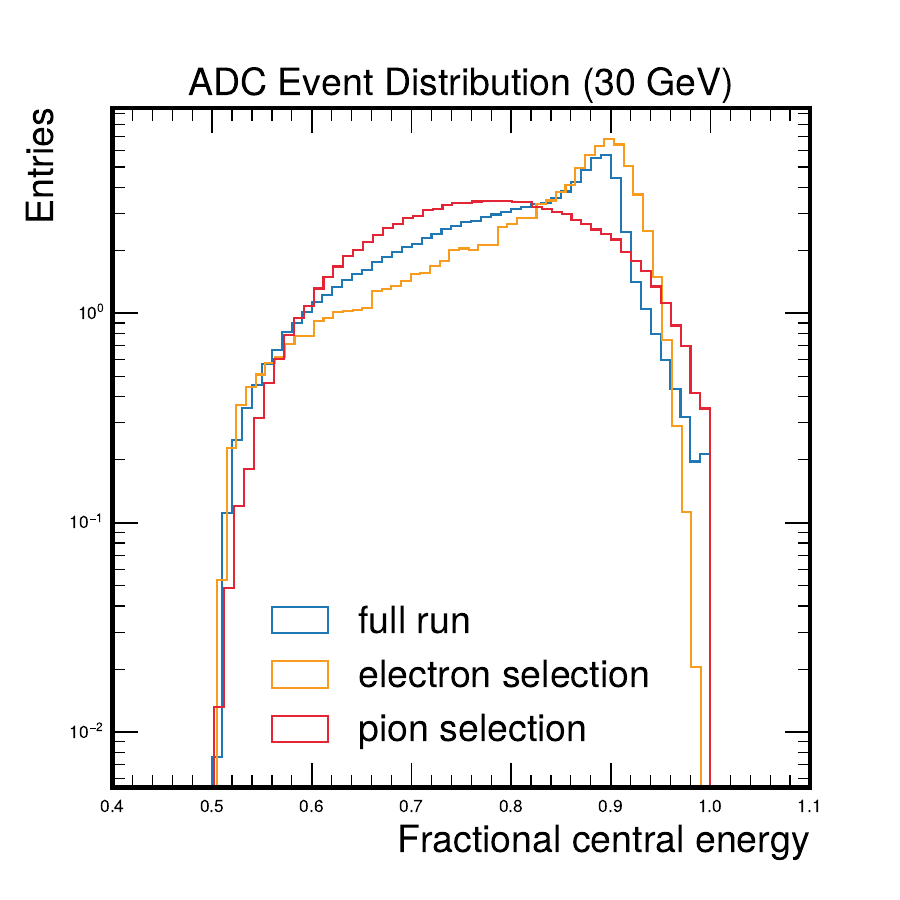}
\caption{Fractional central energy distribution for beam energies of 16, 20 and 30 GeV. The full run, electron region, and pion region distributions are normalized individually.}\label{fig:centralfrac_filter}
\end{figure}

The intensity of the ADC counts in the central channel, with some slight spread in the 4 nearest neighbor channels, suggest that the fractional central energy may adequately quantify the difference between electrons and pions. The fractional central energy is defined as the ADC count of the central channel divided by the summed ADC of the channel plus its 4 nearest neighbors (diagonals excluded). The fractional central energy is plotted per event for the electron and pion regions separately, together with their combinations (Fig.~\ref{fig:centralfrac_filter}). The electron and pion spectra clearly differ, with electrons dominantly peaking around 0.9 across all energies. This seems to suggest the EMCal can detect a narrower shower shape from electrons. However, it is important to note that the pion distribution still has a significant presence in this region, especially in the highest energy runs. 

\begin{figure}[hbtp]
\centering
\includegraphics[width=0.49\textwidth]{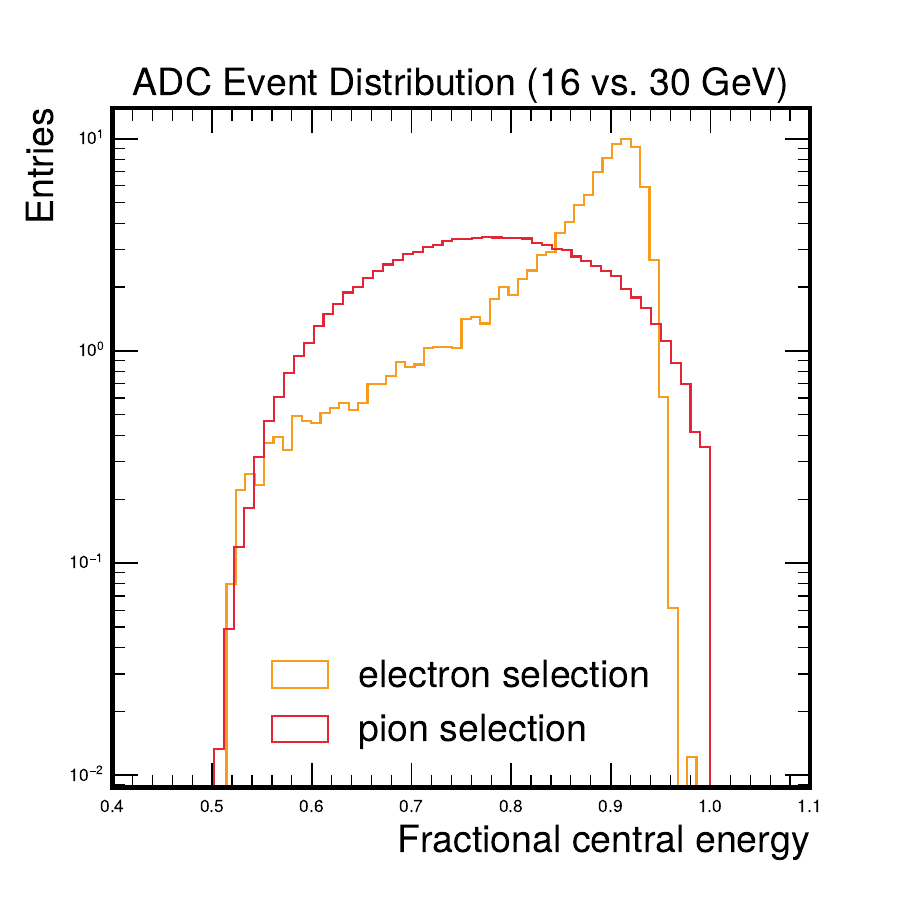}
\includegraphics[width=0.49\textwidth]{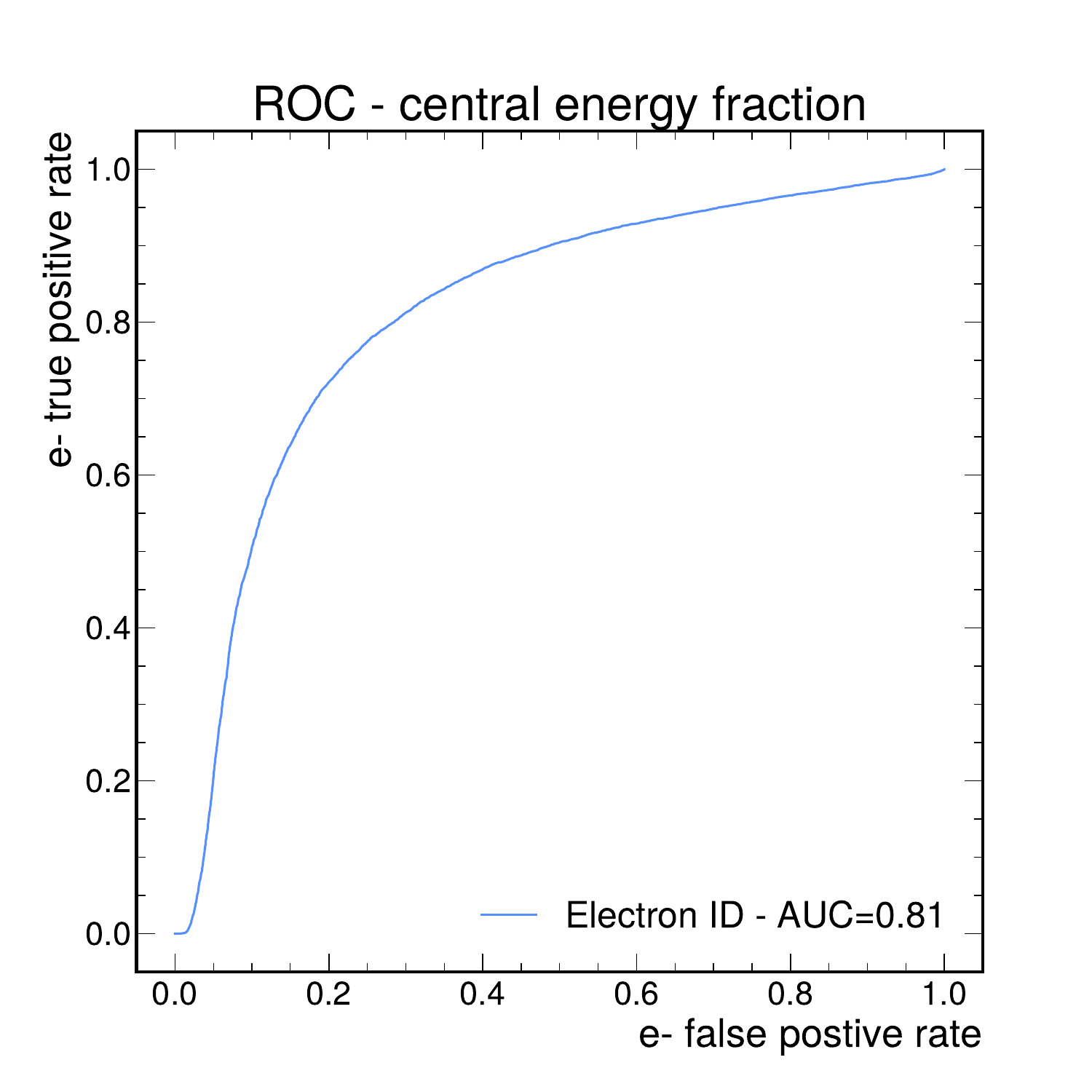}
\caption{Comparison of pion and electron fractional central energy distributions for beam energies of 16 vs. 30 GeV, as well as a ROC curve using the central energy fraction as the discriminator.}\label{peak_comp}
\end{figure}

To confirm that the shape difference is dependent on the particle type and not the energy, two runs at different energies are selected (16 GeV for electrons and 30 GeV for pions), so that the energy deposited in the central module is identical for electrons and pions. The corresponding fractional central energy observable are shown in Fig.~\ref{peak_comp}. The same trend between particles as in Fig.~\ref{fig:centralfrac_filter} is observed, and the dependence of the distribution on particle ID is asserted. Given these findings, the fractional central energy distribution is a candidate to discriminate between the two charged particles. After scanning over the electron and pion spectra of each run, receiver operating characteristic curves (ROC curves) are created, where events that meet the fractional central energy threshold are compared against their respective particle region (the latter roughly corresponding to the truth value).  The performance was found to be consistent across both detector configurations (with attenuator or with neutral density filter) and across all beam energies. The performance was also stable against variations of the pion region selection. This consistency suggests that shower shape is a potential tool for performing particle identification. Fig.~\ref{peak_comp} shows the particle separation with an area under the curve (AUC) over 0.8 for 16 GeV electrons and 30 GeV pions. Additional particle identification in DarkQuest will come from comparisons of the EMCal energy and particle momentum determined from the spectrometer ($E/p$) to achieve the target 99\% pion rejection factor~\cite{Berlin:2018pwi}. A more sophisticated multivariate discriminator (e.g. a neural network) could also be investigated to improve the PID performance of the EMCal. 

\section{Background Rates in DarkQuest's Experimental Cavern}

\begin{figure}[htbp]
    \centering
    \begin{minipage}[b]{0.45\textwidth}
        \centering
        \includegraphics[width=\textwidth]{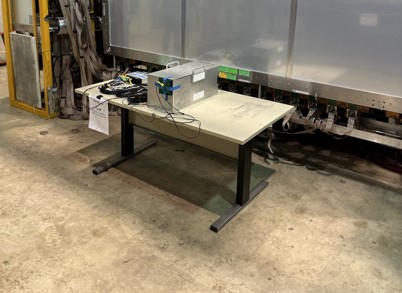}
    \end{minipage}
    \hspace{0.05\textwidth} 
    \begin{minipage}[b]{0.45\textwidth}
        \centering
        \includegraphics[width=\textwidth]{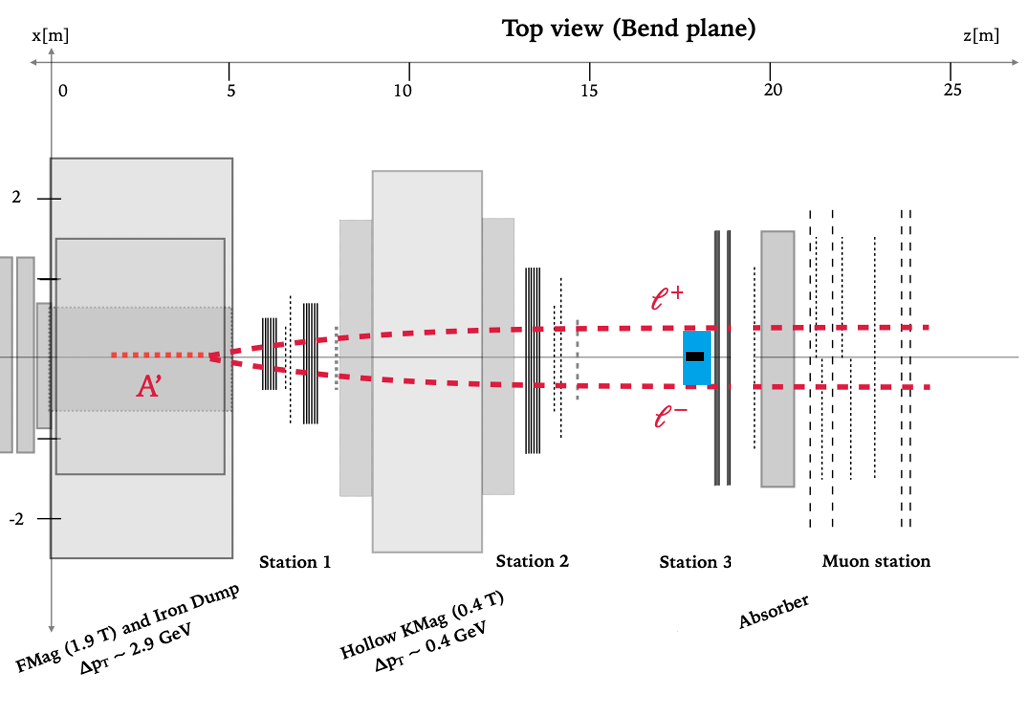}
    \end{minipage}
    \caption{Left: EMCal in the NM4 enclosure. Right: Diagram showing the location of the EMCal within the SpinQuest apparatus. The EMCal is represented in black and the table it stands on is represented in blue. Note that this is a ``top down" view, and while the EMCal is located along the x-axis of the beamline it is not directly centered in the y-axis of the beamline.}
    \label{fig:nm4_location}
\end{figure}

Understanding the background rates in DarkQuest's experimental cavern, NM4, is crucial for the experiment as the EMCal will be used to trigger the readout of the rest of the spectrometer. In order to study these rates, we installed the EMCal in the NM4 Experimental Cavern currently being used by the SpinQuest experiment. As the physical setup and the beam configuration of the SpinQuest experiment is similar to what will be used for DarkQuest, the resulting data will be directly relevant. We installed the EMCal as close to where it would be located for DarkQuest as possible. As shown in Fig.~\ref{fig:nm4_location}, we were able to position it in front of Station 3 in the SpinQuest spectrometer at a low elevation outside the SpinQuest acceptance. The true location for the EMCal installation will be between Station 3 and the Hadron Absorber and centered on the beam line. These studies were performed during SpinQuest runs with a beam intensity of $10^{12}$ protons per spill and with the open-aperture magnet (KMag) off.

\begin{figure}[htbp]
    \centering
    \begin{minipage}[b]{0.46\textwidth}
        \centering
        \includegraphics[width=\textwidth]{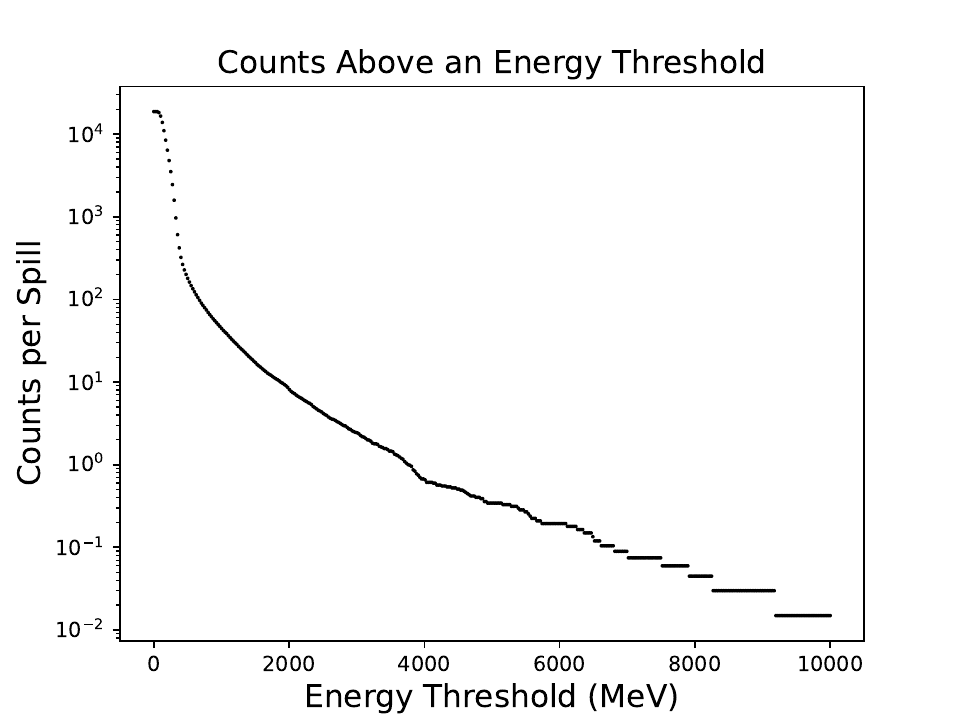}
    \end{minipage}
    \hspace{0.05\textwidth}
    \begin{minipage}[b]{0.46\textwidth}
        \centering
        \includegraphics[width=\textwidth]{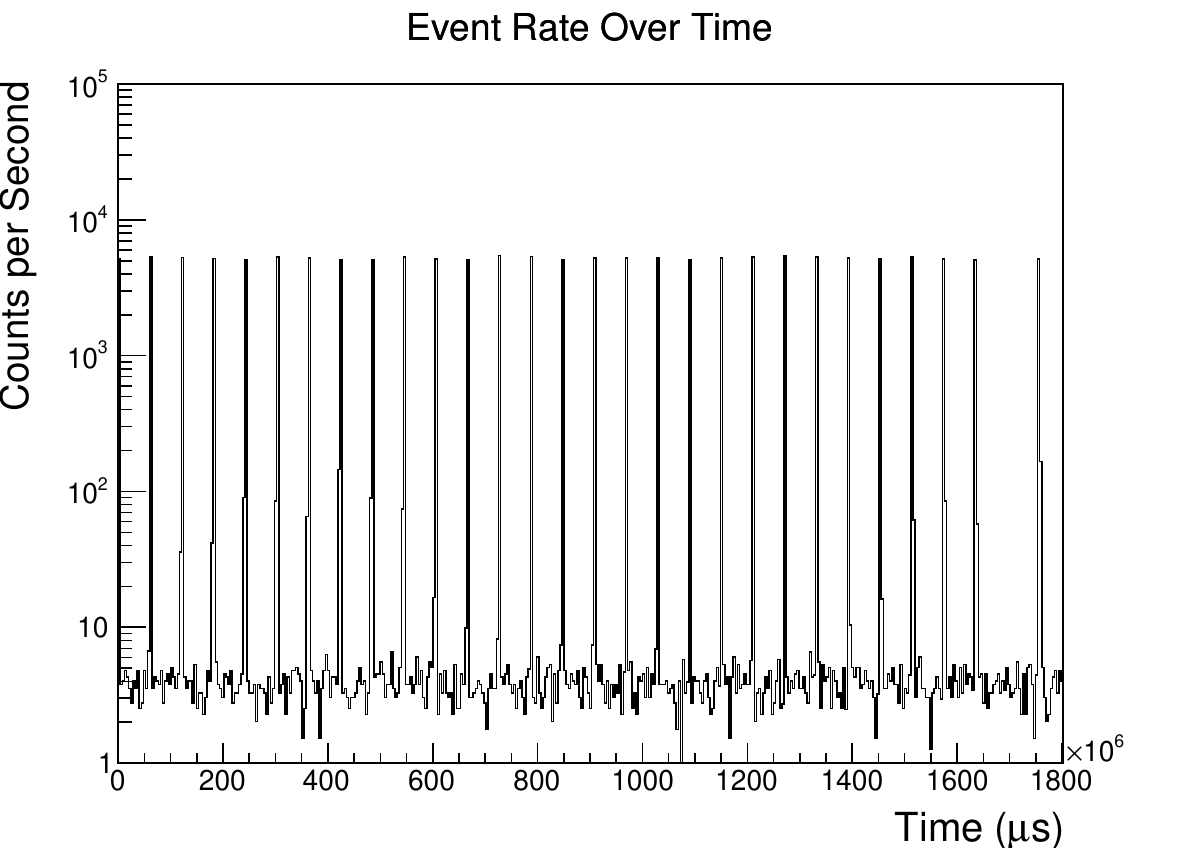}
    \end{minipage}
    \caption{Left: Total number of counts as a function of the threshold on energy for all 4 modules averaged over all spills. The beam intensity was $10^{12}$ particle per spill for the run. Right: Event rate over time for the first 30 minutes of this run, clearly showing intensity spikes from the recurrent beam spills.}
    \label{fig:nm4_background}
\end{figure}

We applied a low energy threshold of 50 MeV to allow for a complete analysis of the event rates throughout the energy spectrum. The resulting  distribution showed little variation between channels and a total event rate of roughly $4\cdot 10 ^{3}$ counts per second, as shown in Fig.~\ref{fig:nm4_background}. We note that 15 of 16 channels were functional during these studies due to physical damage to a cable connection for one of the channels. This event rate, however, was mostly comprised of low energy events. Fig.~\ref{fig:nm4_background} also shows the event rate as a function of energy threshold and demonstrates that the vast majority of these events are less than 1 GeV. This implies that we can increase the energy threshold required for each event and significantly reduce the event rate. We found that the average number of events per spill triggering on an energy greater than 5 GeV is 0.343 and above 10 GeV is 0.015. Although this represents only 4 modules of the 648 total for the EMCal, a naive scaling would lead to an expected background rate of around 55 events per spill above 5 GeV and less than 3 events per spill above 10 GeV for the full-sized EMCal. This is three orders of magnitude below the limitations of the current SpinQuest DAQ system. While the SpinQuest background is not uniform, as this naive scaling would suggest, the extremely low background estimate provides evidence that there are not unanticipated backgrounds (e.g. neutrons) and that the trigger rate can be kept to a sufficiently low level.

\section{Conclusions}
The commissioning of the DarkQuest EMCal was performed with test beam data. The studies on linearity, resolution, and response to various particles confirm the suitability of the EMCal and its newly developed readout electronics for the DarkQuest experiment. Light attenuation with neutral density filters and electrical attenuation were found to give similar performance. The measurements can be used as input to develop an accurate Monte Carlo simulation of the detector and readout electronics. Background rates in the real experiment were also studied and were found to be within the constraints of the existing readout system. 

\section*{Acknowledgments}
The authors would like to thank the teams at the Fermilab Test Beam Facility for their support and contributions to this project. This material is based upon work supported by the National Science Foundation under Grant No. 2320699. ZW was supported by the Universities Research Association’s Visiting Scholars Program. NP was supported by the NSF Graduate Research Fellowship Program. AM was supported by the DOE Office of Science Graduate Student Research Program.



 \bibliographystyle{elsarticle-num} 
 \bibliography{cas-refs}





\end{document}